\documentclass[12pt]{article}

\usepackage{epsfig}

\def\Journal#1#2#3#4{{#1} {\bf #2}, #3 (#4)}

\def\NPB{{\em Nucl. Phys.} B}
\def\PLB{{\em Phys. Lett.}  B}
\def\PRL{\em Phys. Rev. Lett.}
\def\PRD{{\em Phys. Rev.} D}

\def\be{\begin{equation}}
\def\ee{\end{equation}}
\def\bea{\begin{eqnarray}}
\def\eea{\end{eqnarray}}
\newcommand{\lwig}{\mbox{\,\raisebox{.3ex}
    {$<$}$\!\!\!\!\!$\raisebox{-.9ex}{$\sim$}\,}}
\newcommand{\gwig}{\mbox{\,\raisebox{.3ex}
    {$>$}$\!\!\!\!\!$\raisebox{-.9ex}{$\sim$}}\,}

\newcommand{\iai}{I\overline{I}} 
\newcommand{\xpr}{{x^\prime}}

\date{}

\begin{document}
\title{{\normalsize\rightline{DESY 98-061}\rightline{hep-ph/9805492}} 
  \vskip 1cm 
  \bf QCD-Instantons at HERA\thanks{Talk presented at the 
    6th International Workshop on Deep-Inelastic Scattering and QCD (DIS\,98),
    Brussels, April 1998; to be published in the Proceedings 
    (World Scientific).}}
%\vspace{2.5cm}
\author{A. Ringwald and F. Schrempp\\[0.3cm]
Deutsches
  Elektronen-Synchrotron DESY, Hamburg, Germany}

\begin{titlepage} 
\maketitle

\vspace{2cm}
\begin{abstract}
A ``fiducial'' kinematical region for our
calculations of instanton ($I$)-induced processes at HERA within 
$I$-perturbation theory is extracted from recent  
lattice simulations of QCD. Moreover, we present the finalized
$I$-induced cross-sections exhibiting a strongly reduced residual 
dependence on the renormalization scale.
\end{abstract}
\thispagestyle{empty}
\end{titlepage}
\newpage \setcounter{page}{2}

\section{Introduction}\label{s0}
In this contribution, we briefly summarize some recent progress in 
our ongoing systematic study~\cite{rs,grs,dis97-phen,dis97-theo,mrs1,mrs3,crs}
of the discovery potential of DIS events induced by QCD instantons. 

Instantons~\cite{bpst} are non-perturbative gauge field fluctuations. They 
describe {\it tunnelling} transitions between  degenerate ground
states (vacua) of {\it different topology} in non-abelian gauge 
theories like QCD. Correspondingly, (anti-)\-in\-stan\-tons carry an 
{\it integer topologigal} charge $\mid Q\mid =1$, while the usual perturbation 
theory resides in the sector $Q=0$. Unlike the latter, instantons
violate {\it chirality} ($Q_5$) in (massless) QCD and the sum of baryon
plus lepton number ($B+L$) in QFD, in accord~\cite{th} with the general
ABJ-anomaly relation. An experimental discovery of instanton
($I$)-induced  events would clearly be of basic significance.

The deep-inelastic regime is distinguished by the fact that here
hard in\-stan\-ton-induced processes may both be {\it
calculated}\,~\cite{bb,mrs1} within instanton-per\-tur\-ba\-tion
theory and possibly 
{\it detected experimentally}\,~\cite{rs,grs,dis97-phen,crs}.
As a  key feature we have recently shown~\cite{mrs1}, that in DIS the 
generic hard scale ${\cal Q}$ cuts off instantons with {\it large size} 
$\rho>>{\cal Q}^{-1}$, over which we have no control theoretically. 

There has been much recent activity in the lattice community 
to ``measure'' topological fluctuations in 
{\it lattice simulations}~\cite{lattice,ukqcd} of QCD. Being
independent of perturbation theory, such simulations provide
``snapshots'' of the QCD vacuum including all possible
non-perturbative features like instantons. They also provide crucial 
support for important prerequisites of our calculations in DIS,
like the validity of $I$-perturbation theory and 
the dilute $I$-gas approximation for {\it small} instantons of size
$\rho \leq {\cal Q}^{-1}$.  As one main point of  this paper (Sect. 2), 
these lattice constraints will be exploited and translated into a 
``fiducial'' kinematical region for our predictions of the
instanton-induced DIS cross-section based on $I$-perturbation theory. 
In Sect. 3 we display the finalized calculations of the various
instanton-induced cross-sections~\cite{mrs3}. The essential new 
aspect here is the
strong reduction of the residual dependence on the renormalization
scale $\mu_r$ resulting from a recalculation based on improved instanton
densities~\cite{morretal}, which are renormalization group (RG)
invariant at the 2-loop level.
\section{Validity of Instanton Perturbation Theory in DIS\\
       -- Restrictions from Lattice-QCD Simulations}\label{s1}

The leading instanton (I)-induced process in the DIS regime of $e^\pm
P$ scattering is displayed in Fig.\,1. The non-trivial {\it topology}
of instantons is reflected in a  violation of {\it chirality} by 
$\mid\Delta Q_5\mid =2 n_f$, in accord~\cite{th} with the general ABJ-anomaly 
relation (while in pQCD always $\Delta Q_5\equiv 0$).
The dashed box emphasizes the  so-called  instanton-{\it subprocess}
with its own Bjorken variables,
        \begin{equation}
        Q^{\prime\,2}=-q^{\prime\,2}>0;\hspace{0.5cm}
        0\le\xpr=\frac{Q^{\prime\,2}}{2 p\cdot q^\prime}\le 1.
        \end{equation}         
%%%%%%%%%%%%%%%%%%%%%%%%%%%%%FIGURE  %%%%%%%%%%%%%%%%%%%%%%%%%%%%%%%%%
\begin{figure}
\vspace{-0.8cm}
\begin{center}
\epsfig{file=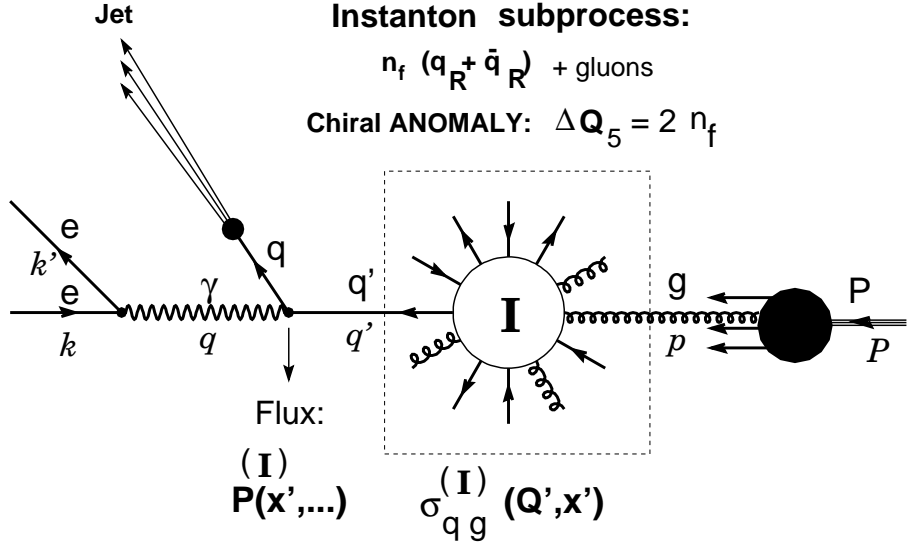,width=12cm}
\caption[dum]{The leading instanton-induced process 
in the DIS regime of $e^\pm P$ scattering, violating {\it chirality} 
by $\mid\Delta Q_5\mid =2 n_f$.}
\end{center}
\end{figure}
%%%%%%%%%%%%%%%%%%%%%%%%%%%%%%%%%%%%%%%%%%%%%%%%%%%%%%%%%%%%%%%%%%%%%% 
The cross-section of interest may be 
shown~\cite{rs,dis97-phen,mrs3} to exhibit a
convolution-type structure, consisting of a smooth, calculable 
``flux factor'' $P^{(I)}(\xpr,\ldots)$ from the
the $\gamma^\ast \bar{q}q\prime$ vertex, and the $I$-subprocess
total cross-section $\sigma^{(I)}_{q^\prime\,g}(Q^\prime,\xpr)$, 
containing the essential instanton dynamics. We have 
evaluated the latter~\cite{mrs3} 
by means of the optical theorem and the so-called  
$\iai$-valley approximation~\cite{valley} for the relevant 
$q^\prime g\Rightarrow q^\prime g$ forward elastic scattering
amplitude in the $\iai$ background. This method 
resums the exponentiating final state gluons in form of the known
valley action $S^{\iai}$ and reproduces standard $I$-perturbation theory at
larger  $\iai$ separation $\sqrt{R^2}$. 

Corresponding to the symmetries of the theory, the instanton calculus
introduces at the classical level certain (undetermined) 
 ``collective coordinates'' like the $I\,(\overline{I})$-size parameters 
$\rho\,(\overline{\rho})$ and the $\iai$ distance 
$\sqrt{R^2/\rho\overline{\rho}}$ (in units of the size).  
Observables like $\sigma^{(I)}_{q^\prime\,g}(Q^\prime,\xpr)$,
must be independent thereof and thus involve integrations over all collective
coordinates. Hence, denoting the {\it density} of $I\,(\overline{I})$'s by
$D(\rho (\overline{\rho}))$ (see Eq.\,(\ref{density})), we have generically,
  \begin{equation}
        \begin{array}{lcl}
        \sigma^{(I)}_{q^\prime g}( Q^\prime,x^\prime)
        &=&\int\limits_0^\infty {\rm d}\rho
        \int\limits_0^\infty {\rm d}\overline{\rho}\underbrace{D(
        \rho)D(\overline{\rho})}_
        {I,\overline{I}-{\rm densities}
        \Leftarrow {\bf \rm  Lattice!}} 
        e^{-({\rho + \overline{\rho}})Q^\prime}\\
        &\times&\int\limits^\infty {\rm d}\xi
        {\mathcal M}(\xi, x^\prime, Q^\prime,\ldots)
        e^{-\frac{4\pi}{\alpha_s}(S^{\iai}(\xi)-1)},\\
        \end{array}\label{sigma}
  \end{equation}   
where
$\xi=R^2/\rho\overline{\rho}+\rho/\overline{\rho}+\overline{\rho}/\rho$ 
is a convenient conformally invariant variable characterizing the 
$\iai$ distance. In Eq.(\ref{sigma}), the crucial exponential 
cut-off~\cite{mrs1}
$e^{-(\rho+\overline{\rho})\,Q^\prime}$ is responsible for the {\it
finiteness} of the $\rho,\,\overline{\rho}$ integrations. In addition,
it causes the integrals (\ref{sigma}) to be dominated by a {\it
single, calculable} (saddle) point
($\rho^\ast=\overline{\rho}^\ast\sim Q^{\prime\,-1},
\xi^\ast(\xpr,Q^\prime)$), in {\it one-to-one}
relation to the conjugate momentum variables ($\xpr, Q^\prime$). 
This effective one-to-one
mapping of the conjugate $I$-variables allows for the following 
important strategy:
We may determine {\it quantitatively} the range of validity
of $I$-perturbation theory and the dilute $I$-gas approximation in the 
instanton collective coordinates ($\rho<\rho_{\rm max},
R/\rho>(R/\rho)_{\rm min}$) from recent (non-perturbative) lattice
simulations of QCD and translate the resulting constraints via the
mentioned one-to-one relations into a
``fiducial'' kinematical region ($\xpr>\xpr_{\rm min},Q^\prime>
Q^\prime_{\rm min}$) at HERA! Experimentally, these cuts must be
implemented via a ($\xpr,\,Q^\prime$) reconstruction from the final
state topology~\cite{crs}, while theoretically, they are incorporated
into our $I$-event generator~\cite{grs} ``QCDINS 1.6.0'' and the
resulting prediction of
$\sigma^{(I)}_{\rm HERA}(\xpr>\xpr_{\rm min},Q^\prime>Q^\prime_{\rm
min})$ (see Sect.\,3).  

In lattice simulations 4d-Euclidean space-time is made discrete,
e.g. in case of the ``data'' from the UKQCD
collaboration~\cite{ukqcd} which we shall use here,  
\begin{center}
\begin{tabular}{lcl}
lattice spacing: a &=& 0.055 - 0.1 fm\\
lattice volume:  V&=&$l_{\rm space}^{\,3}\cdot l_{\rm time}=
[16^3\cdot 48 - 32^3\cdot 64]\,a^4$\\
\end{tabular}
\end{center}
In principle, such a lattice allows to study the properties of an 
ensemble of (anti-)instantons  with sizes $a < \rho < V^{1/4}$. However,
in order to make instanton effects visible, a certain ``cooling'' 
procedure has to be applied first. It is designed to
filter out (dominating) fluctuations of {\it short} wavelength
${\cal O}(a)$, while affecting the instanton fluctuations of much longer
wavelength $\rho >> a$ comparatively little. For a discussion of 
lattice-specific caveats, like possible lattice artefacts and the
dependence of results on ``cooling'' etc., see Refs.~\cite{lattice,ukqcd}.   

%%%%%%%%%%%%%%%%%%%%%%%%%%%%%FIGURE  %%%%%%%%%%%%%%%%%%%%%%%%%%%%%%%%%%
\begin{figure}
\begin{center}
\epsfig{file=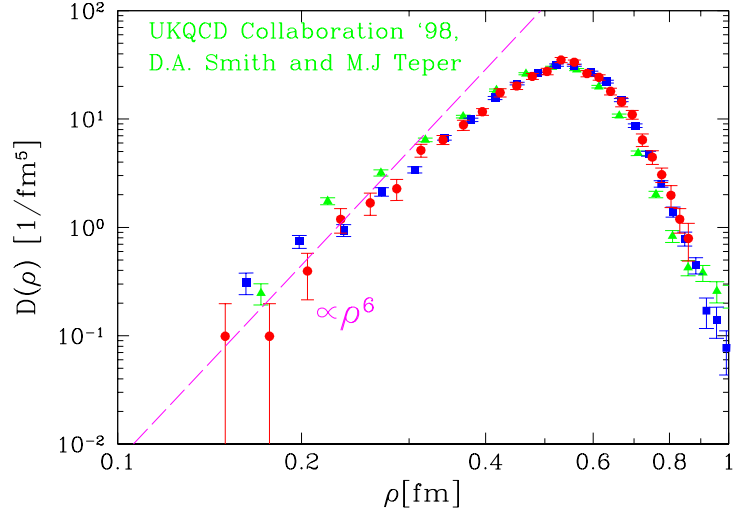,angle=-90,width=11cm}
\epsfig{file=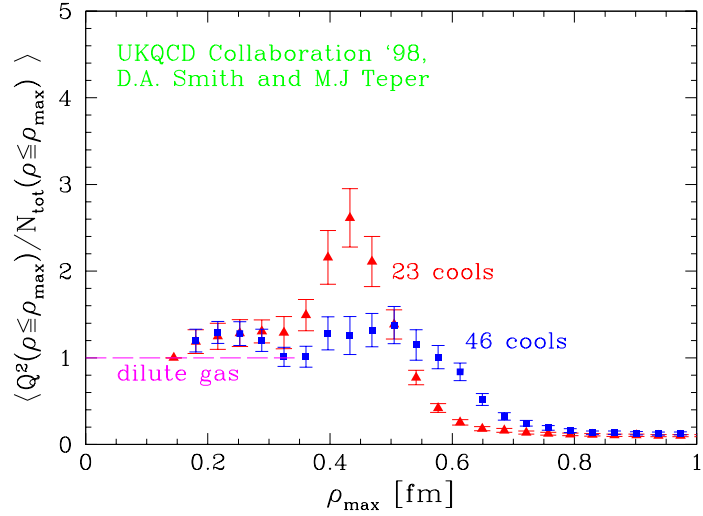,angle=-90,width=11cm}
\caption[dum]{Support for the validity of $I$-perturbation theory for the
$I$-density $D(\rho)$  (top) and the dilute $I$-gas approximation
(bottom) for $\rho<\rho_{\rm max}\simeq 0.3$ fm from recent lattice
data~\cite{ukqcd}.}
\end{center}
\end{figure}
%%%%%%%%%%%%%%%%%%%%%%%%%%%%%%%%%%%%%%%%%%%%%%%%%%%%%%%%%%%%%%%%%%%%%%
The first important quantity of interest, entering Eq.\,(\ref{sigma}), is 
the $I$-density, $D(\rho)$ (tunnelling probability!).
It has been worked out a long time ago in the framework of 
$I$-perturbation theory: (renormalization scale $\mu_r$)
\begin{equation}
   D(\rho)\equiv \frac{{\rm d}n}{{\rm d}^4 x {\rm d}\rho}=
   d \left(\frac{2\pi}{\alpha_s(\mu_r)}\right)^6
   \exp{(-\frac{2\pi}{\alpha_s(\mu_r)})}\frac{(\rho\, \mu_r)^b}{\rho^{\,5}}.
   \label{density} 
\end{equation}
Note the {\it power law} in the instanton size $\rho$ with the
power $b$ given in Table\,1,
\begin{table}[h]
\vspace{-0.3cm}
\caption[dum]{The power $b$ in Eq.\,(\ref{density}) from 
Ref.~\cite{th} and Ref.~\cite{morretal}, making the
$I$-density $D(\rho)$ RG-group invariant at the 1-loop and
2-loop level, respectively.}
\begin{center}
\footnotesize
      \begin{tabular}{|c|c|c|}
      \hline
     $b$&$\rule[-3mm]{0mm}{7mm}\frac{1}{D}\frac{{\rm d}D}{{\rm d}\mu_r}$&Ref.\\
      \hline
      $\beta_0$& ${\mathcal O}(\alpha_s)$& {'t Hooft~\cite{th}}\\
      \hline
      &&Morris, Ross,\\
      \raisebox{1.5ex}[-1.5ex]{$\beta_0+\frac{\alpha_s(\mu_r)}{4\pi}
      (\beta_1-12\beta_0)$} &\raisebox{1.5ex}[-1.5ex]{
      ${\mathcal O}(\alpha_s^2)$}&\& Sachrajda~\cite{morretal}\\\hline 
      \end{tabular}
\end{center}
\end{table}
\noindent
in terms of the QCD $\beta$-function coefficients:
$\beta_0=11-\frac{2}{3}{n_f};\ \beta_1=102-\frac{38}{3} {n_f}$.
This power law $D(\rho)_{\mid n_f=0}\propto \rho^6$ of
$I$-perturbation theory is confronted in Fig.\,2\,(top) with recent 
lattice ``data'',
which strongly suggests $I$-perturbation theory to be valid for $\rho \lwig 
\rho_{\rm max}=0.3$ fm. 
Next, consider the square of the total topological charge,
$Q^2=(n\cdot(+1)+\bar{n}\cdot(-1))^2$ along with the
total number of charges $N_{\rm tot}=n+\bar{n}$.
For a {\it dilute gas}, the number fluctuations are {\it poissonian} and 
correlations among the $n$ and $\bar{n}$ distributions
absent, implying $\langle (n-\bar{n})^2\rangle = \langle n+\bar{n}\rangle$,
or $\langle \frac{Q^2}{N_{\rm tot}} \rangle =1$.
From Fig.\,2\,(bottom), it is apparent that this relation
characterizing the validity of the dilute $I$-gas approximation, is well 
satisfied for sufficiently {\it small} instantons! Again, we find 
$\rho_{\rm max}\simeq 0.3$ fm, quite independent of the number of
cooling sweeps. For increasing  $\rho_{\rm max}\gwig 0.3$ fm, the ratio 
$\langle \frac{Q^2}{N_{\rm tot}}\rangle$ rapidly and strongly deviates
from one.
Crucial information about a second instanton 
parameter of interest, the average $\iai$ distance $ \langle R \rangle$, may be
obtained as well from the lattice \cite{lattice,ukqcd}. 
Actually, the ratio~\cite{ukqcd} 
  $ \frac{\langle R_{\iai}\rangle}{\langle \rho \rangle}\simeq 0.83$
has good stability against ``cooling'', from which we shall take
$R/\rho\gwig 1$ as a reasonable lower limit for our $I$-perturbative 
DIS calculations.  

Finally, the ``fiducial'' kinematical region for our cross-section
predictions in DIS  is found from lattice constraints and the
discussed saddle-point translation as
\begin{equation}
 \left.\begin{array}{lcl}\rho&\lwig& 0.3 {\rm\ fm};\\
 \frac{R}{\rho}&\gwig&1\\
 \end{array}\right\}\Rightarrow
 \left\{\begin{array}{lclcl}Q^\prime&\gwig&Q^\prime_{\rm min}&\simeq&
 8 {\rm\ GeV};\\
 x^\prime&\gwig&x^\prime_{\rm min}(Q^\prime_{\rm min})&\simeq &0.35.\\
 \end{array} \right .
\label{fiducial}
\end{equation}
\section{$I$-Induced Cross-Sections for HERA}\label{s2}

We have achieved great progress in stability by using the 
{\it 2-loop RG invariant} form of the $I$-density $D(\rho)$ from 
Eq.\,(\ref{density}) and Table\,1 in a recalculation of the 
$I$-subprocess cross-sections~\cite{mrs3}: The  
residual dependence on the renormalization scale $\mu_r$ turns out to
be {\it strongly reduced} (Fig.\,3).
As ``best'' scheme we use $\mu_r = 0.15\ Q^\prime$ throughout, for which 
$\partial \sigma^{(I)}_{q^\prime g}/\partial \mu_r \simeq 0$. 
The quantitative calculations of $\sigma^{(I)}_{\rm q^\prime g}$ 
%%%%%%%%%%%%%%%%%%%%%%%%%%%%%FIGURE  %%%%%%%%%%%%%%%%%%%%%%%%%%%%%%%%%%%
\begin{figure}
\begin{center}
\epsfig{file=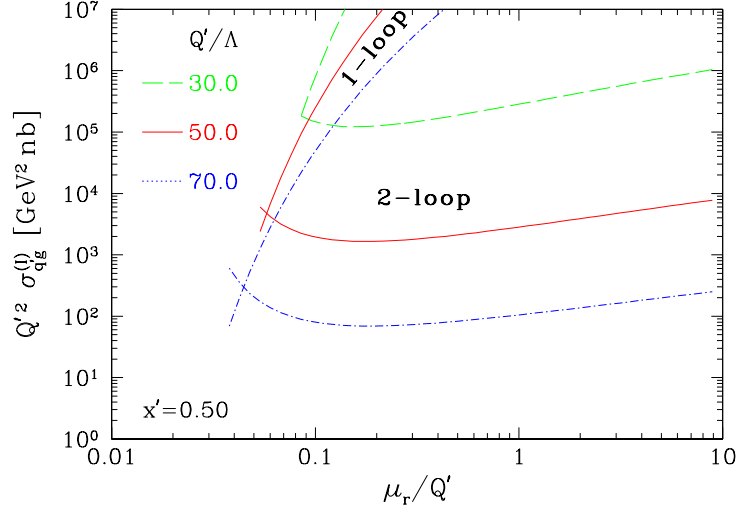,angle=-90,width=11cm}
\caption[dum]{Renormalization-scale dependence of the $I$-subprocess 
cross-section}
\end{center}
\end{figure}
%%%%%%%%%%%%%%%%%%%%%%%%%%%%%%%%%%%%%%%%%%%%%%%%%%%%%%%%%%%%%%%%%%%%%%
(Fig.\,4) nicely illustrate the qualitative arguments from 
Sect.\,2, that the $Q^\prime$ dependence  probes 
the effective instanton {\it size $\rho$} (top), 
while the $x^\prime$ dependence maps the 
$\iai$ {\it distance $R$} in units of the $I$-size $\rho$ (bottom).
%%%%%%%%%%%%%%%%%%%%%%%%%%%%%FIGURE  %%%%%%%%%%%%%%%%%%%%%%%%%%%%%%%%%%%5
\begin{figure}[h]
\begin{center}
\epsfig{file=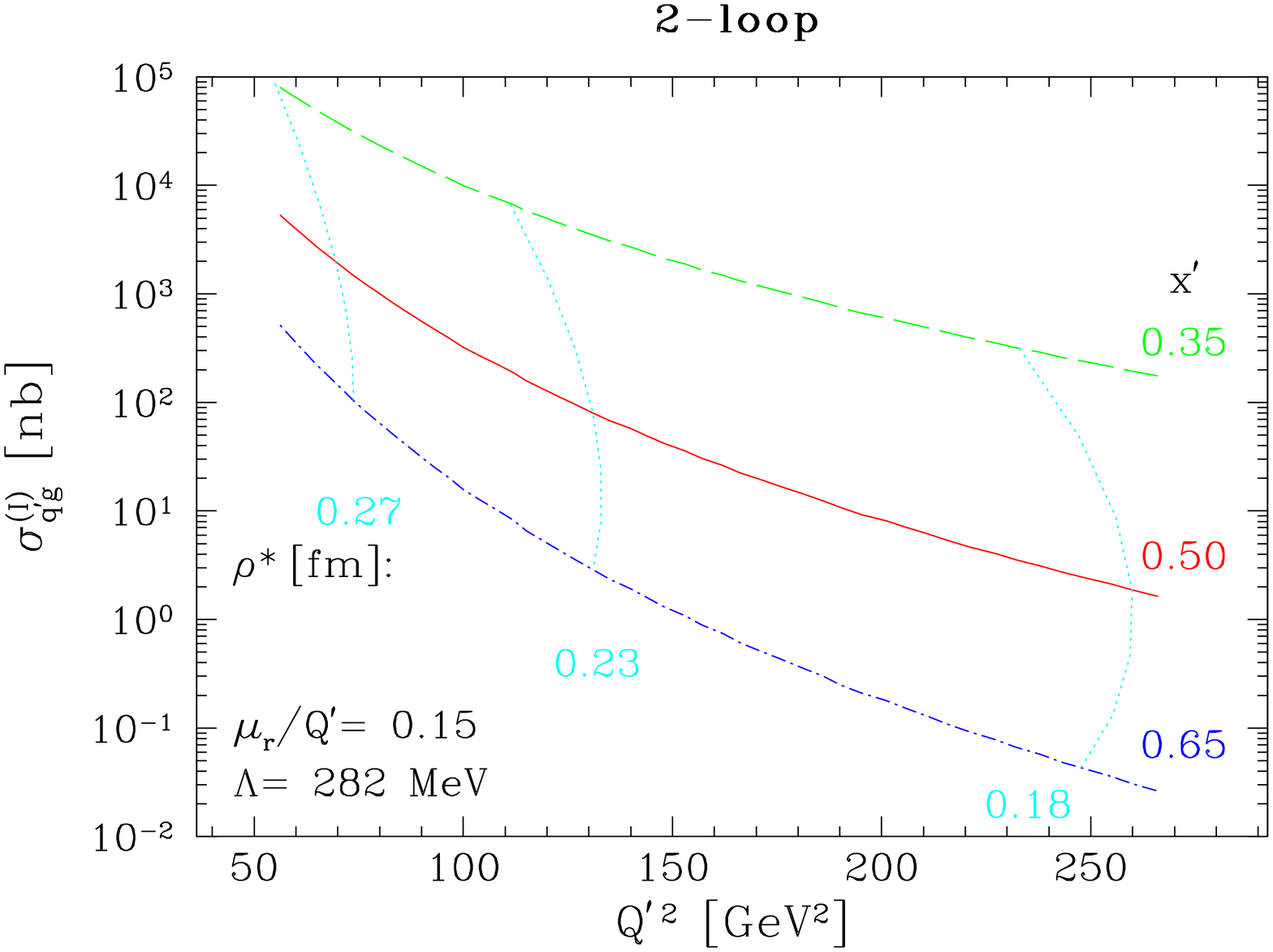,angle=-90,width=9.1cm}
\epsfig{file=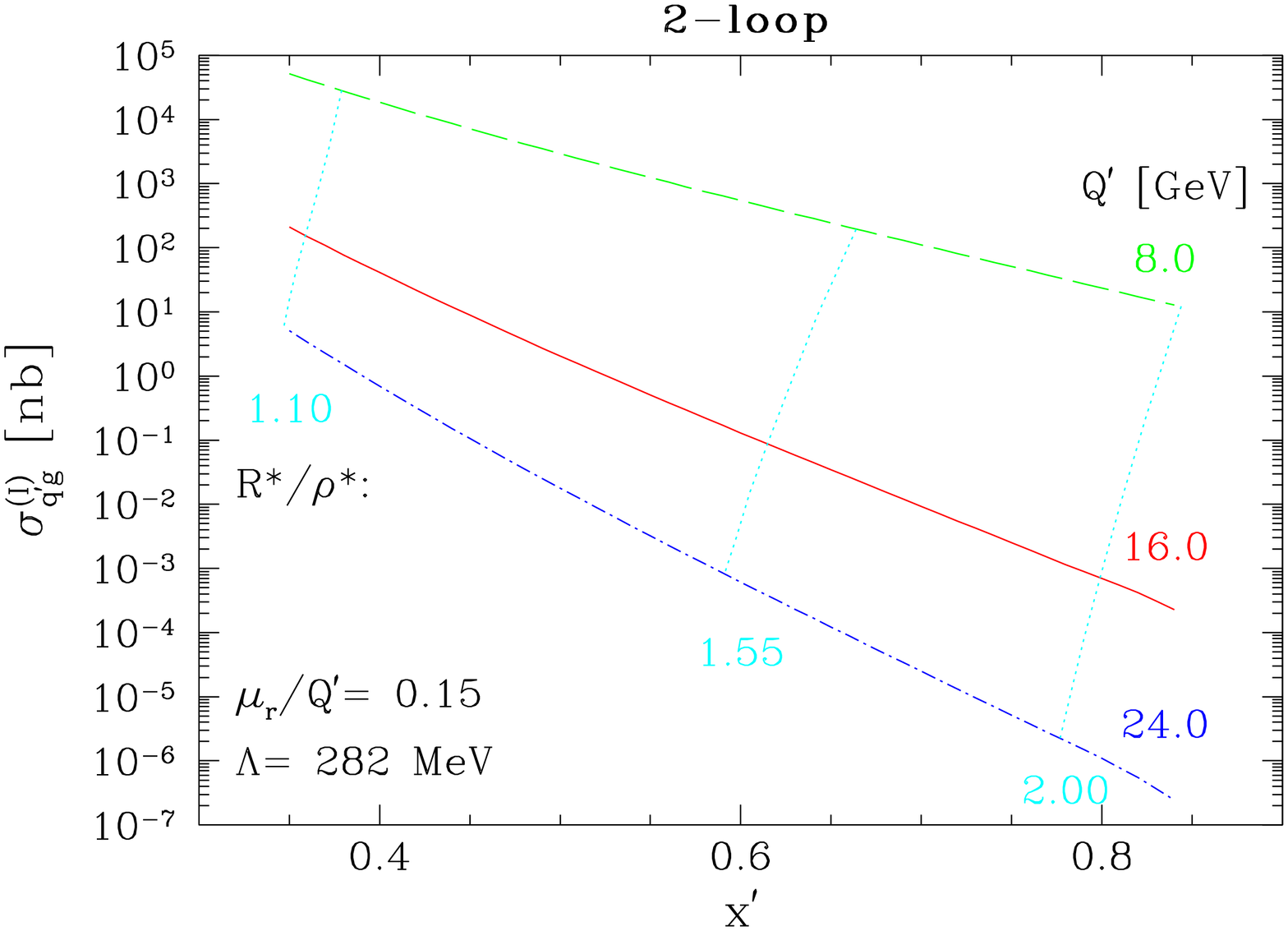,angle=-90,width=9.1cm}
\caption[dum]{Instanton-induced cross-sections}
\end{center}
\end{figure}
%%%%%%%%%%%%%%%%%%%%%%%%%%%%%%%%%%%%%%%%%%%%%%%%%%%%%%%%%%%%%%%%%%%%%%

Fig.\,5 displays the finalized $I$-induced cross-section at HERA,
as function of the cuts $x^\prime_{\rm min}$ and
$Q^\prime_{\rm min}$ in leading semi-classical approximation, as
obtained with the new release ``QCDINS 1.6.0'' of our $I$-event
generator. 
%%%%%%%%%%%%%%%%%%%%%%%%%%%%%FIGURE  %%%%%%%%%%%%%%%%%%%%%%%%%%%%%%%%%%%5
\begin{figure}[ht]
\begin{center}
\epsfig{file=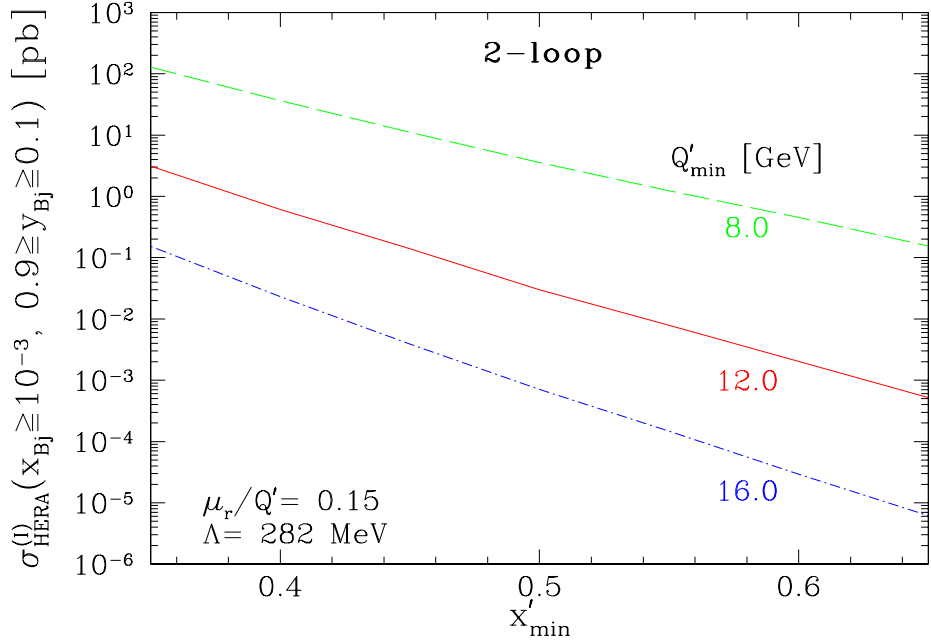,angle=-90,width=14cm}
\caption[dum]{Instanton-induced cross-section at HERA}
\end{center}
\end{figure}
%%%%%%%%%%%%%%%%%%%%%%%%%%%%%%%%%%%%%%%%%%%%%%%%%%%%%%%%%%%%%%%%%%%%%%
For the minimal cuts 
(\ref{fiducial}) extracted from lattice simulations, we
specifically obtain 
\begin{equation}
\sigma^{(I)}_{\rm HERA}(\xpr\ge0.35,Q^\prime\ge 8\, {\rm GeV}) 
\simeq 126\, {\rm pb};
\ x_{\rm Bj}\ge 10^{-3};\ 0.9\ge y_{\rm Bj}\ge 0.1 .
\end{equation}
In view of the fact that the cross-section varies strongly as a
function of the ($\xpr,Q^\prime$) cuts, the constraints from lattice
simulations are extremely valuable for making concrete predictions.

\end{document}